\documentclass[aps,prl,twocolumn,showpacs,preprintnumbers,amsmath,amssymb]{revtex4}
\usepackage{bm}

\begin{document}

\title{Lockin to Weak Ferromagnetism in
TbNi$_2$B$_2$C and ErNi$_2$B$_2$C}
\author{M. B. Walker}
\affiliation{Department of Physics,University of Toronto,Toronto, Ont. M5S 1A7 }
\author{C. Detlefs}
\affiliation{European Synchrotron Radiation Facility, Bo\^ite Postale 220,
    38043 Grenoble Cedex, France}
\date{\today}

\begin{abstract}
This article describes a model in which ferromagnetism necessarily
accompanies a spin-density-wave lockin transition in the
borocarbide structure provided the commensurate phase wave vector
satisfies $Q = (m/n)a^\ast$ with $m$ even and $n$ odd. The results
account for the magnetic properties of TbNi$_2$B$_2$C, and are
also possibly relevant also for those of ErNi$_2$B$_2$C.
\end{abstract}

\pacs{PACS numbers: }

\maketitle

\section{Introduction}
The material TbNi$_2$B$_2$C is one of a number of rare earth
borocarbide materials that display a fascinating variety of
magnetic and superconducting phases.\cite{mul01} This particular
material has a phase transition to a spin-density-wave phase at
$T_N \approx$ 15 K followed by the appearance of weak
ferromagnetism at $T_{WFM} \approx $ 8 K.\cite{cho96} A neutron
diffraction study \cite{der96} of the magnetic structure showed
that the wave vector of the spin-density wave decreased with
decreasing temperature in the spin-density wave phase until the
transition temperature to the weak ferromagnetic phase was
reached, at which point the spin-density-wave wave vector became
approximately constant at $\mathbf{Q} = .545 \mathbf{a}^\ast$.
This behavior is characteristic of a continuous lockin transition
to a commensurate phase occurring at $T_{WFM}$, the commensurate
wave vector being $\mathbf{Q} = .545 \mathbf{a}^\ast = (6/11)
\mathbf{a}^\ast$. This value for the locked in wave vector has
been confirmed by recent high resolution magnetic X-ray scattering
studies.\cite{det02} There is some hysteresis observed at
$T_{WFM}$, which is consistent with the idea that hysteresis can
occur at a continuous commensurate-incommensurate transition
(as well as at a first order commensurate-incommensurate
transition).

Similarly, in ErNi$_2$B$_2$C, there is a transition to a spin density wave phase,
followed apparently by a transition to a weak ferromagnetic phase
\cite{can96} as the temperature is lowered.\cite{sin95,det99,cho01,kaw02}
The study of the weak ferromagnetism is made more complicated
in this case because the magnetic transitions occur
within a superconducting phase. Here also a lockin phase transition appears
to accompany the transition to weak ferromagnetism. (The evidence for the
lockin is that the spin-density-wave wave vector appears to become more or
less independent of temperature below $T_{WFM}$).  The lockin wave vector is
close to $\mathbf{Q} = 0.548 \mathbf{a}^\ast = (17/31) \mathbf{a}^\ast$
according to \cite{cho01} and close to
$\mathbf{Q} = 0.550 \mathbf{a}^\ast = (11/20) \mathbf{a}^\ast$ according
to \cite{kaw02,det02}.

A curious aspect of the behavior of both TbNi$_2$B$_2$C and
ErNi$_2$B$_2$C is that the apparent incommensurate-commensurate
transition and the transition to weak ferromagnetism seem to occur
at the same temperature.  It is possible for two independent
second-order phase transitions to occur at nearly the same
temperature if the interactions in the system happen to have just
the right values so that this happens. However, this is very
unusual, and to have it happen in two different materials is even
more unusual. This suggests that the lockin transition and the
transition to weak ferromagnetism are not independent, but have a
common origin. The purpose of this article is to describe a model
in which  weak ferromagnetism necessarily occurs simultaneously
with a lockin transition in a spin-density-wave state, and that
could therefore be relevant to the magnetic behavior of
TbNi$_2$B$_2$C and ErNi$_2$B$_2$C. The model described here is
related to that proposed in \cite{zhu86}.

One of the results derived below is that weak ferromagnetism
necessarily accompanies the lockin transition in the spin-density
wave states of TbNi$_2$B$_2$C and ErNi$_2$B$_2$C only if the
commensurate wave vector is of the form ${\mathbf Q} = (m/n)
{\mathbf a}^\ast$, with $m$ an even integer and $n$ an odd
integer.  An independent microscopic calculation of the nature of
commensurate phases of ErNi$_2$B$_2$C, including the possible
occurrence of ferromagnetism,  has been carried out in
\cite{jen02}for a number of different commensurate wave vectors.
These results for specific wave vectors are consistent with the
general rule concerning the occurrence of ferromagnetism developed
in this article.

The work described below makes use of a Ginzburg-Landau type of
analysis. This is a useful way of obtaining results of general
validity which depend on the symmetry of the problem.  Thus, the
approach is complementary to the microscopic type of calculation
described in \cite{jen02}.

\section{The Spin-Density-Wave Phase}
A model specifically related to TbNi$_2$B$_2$C will now be described.
Above the spin-density-wave transition temperature the
material TbNi$_2$B$_2$C has a body-centered tetragonal structure
with space-group symmetry I4/mmm.
The spin-density wave in TbNi$_2$B$_2$C is assumed to be
represented by the equation
\begin{equation}
\mathbf {S(r)} = \sum_{\mathbf R, \tilde{B},n,i} \left[ \mathbf{S}_{n,i}
e^{in\mathbf{Q_i \cdot r}} +
\mathbf{S}_{n,i}^\ast e^{-in\mathbf{Q}_i\cdot r} \right]
\delta(\mathbf{r - R - \tilde{B}}). \label{SDW}
\end{equation}
Here ${\mathbf R}$ labels the unit cell,
$\mathbf {\tilde{B}} = 0, \tfrac{1}{2} (\mathbf{ a + b + c})$ labels the magnetic
ions in the crystallographic unit cell (at the corner and body center
positions), n labels the harmonic,  and $i = a, b$ labels the symmetry-equivalent wave vectors
$\mathbf{Q}_a = (Q,0,0)$ and
$\mathbf{Q}_b = (0,Q,0)$, where
$Q \approx 0.55 a^\ast$. Since the spin-density wave in TbNi$_2$B$_2$C
is thought to be longitudinally polarized\cite{lyn97},
i.e. $\mathbf{S}_i$ is parallel
to $\mathbf{Q}_i$ (see however
\cite{kre02}), the nonzero components of the primary order parameter
 $\mathbf{S}_{1,i}$  (describing the first harmonic) are
$S_{1,a,x}$ and $S_{1,b,y}$, which will be denoted below simply by $S_x$ and
$S_y$, respectively.  It is important to remember, however, that $S_x$
and $S_y$ are complex numbers giving the amplitude and phase of the
spin-density waves with wave vectors $\mathbf{Q}_a$ and $\mathbf{Q}_b$,
respectively. (A similar model relevant to ErNi$_2$B$_2$C would involve
$S_{1,a,y}$ and $S_{1,b,x}$ since the spin-density wave in this material is
transversely polarized.\cite{sin95})

Given the above description of the spin-density wave, the
Landau free energy describing a second-order phase transition
to the spin-density-wave phase has the form
\begin{eqnarray}
F & =& A(|S_x|^2 + |S_y|^2)  \nonumber \\
    & & + B(|S_x|^2 + |S_y|^2)^2
    +C|S_x|^2|S_y|^2,
\label{FSDW}
\end{eqnarray}
where $A \propto (T - T_N)$, with $T_N$ being the transition
temperature.  The constants $B$ and $C$ must satisfy $B>0$ and
$4B+C > 0$ for stability.  If $C < 0$ a tetragonal double-Q
spin-density wave phase with $|S_x| = |S_y| \neq 0$ occurs, while
if $C>0$ an orthorhombic single-Q phase with either $|S_x| \neq 0$
or $|S_y| \neq 0$, but not both, occurs. Below T$_N$ in
TbNi$_2$B$_2$C, the structure becomes
orthorhombic\cite{kre01,det02} so it is clearly the single-Q state
that occurs in this material. A given material normally contains both
single-Q domains (in spatially different regions of the crystal).  In
what follows we study explicitly only the properties of the single-Q
domain characterized by a non-zero $S_x$; the properties of the
domain characterized by $S_y$ follow immediately by rotating by
$\pi/2$ about the $c$-axis.

Spin-density-wave harmonics of order $2n+1$ with $n$ an integer are
induced by contributions to the free energy proportional to
$S_{2n+1,x}(S_x^\ast)^{2n+1} + S_{2n+1,x}^\ast S_x^{2n+1}$. Since
this term is linear in $S_{2n+1,x}$, the value of $S_{2n+1,x}$ which
minimizes the free energy must be non-zero. There are other terms in
the free energy that are also linear in $S_{2n+1,x}$ and which also
contribute to the induction of the $(2n+1)^{th}$ harmonic. However,
since we are only interested here in demonstrating that the symmetry
of the problem requires the existence of the $(2n+1)^{th}$ harmonic
(and do not attempt to evaluate its magnitude) it is sufficient to
consider only one example of the terms linear in $S_{2n+1,x}$.
Similarly, throughout this article, only one example of the type of
term necessary for our purpose will be given. The
wave vector associated with $S_{2n+1,x}$ is $(2n+1)\mathbf{Q}_a$.
Similarly, charge-density-wave (or, equivalently, longitudinal
lattice-displacement-wave) harmonics of order $2n$ and complex
charge-density-wave amplitude $\rho_{2n}$ are induced by terms
in the free energy
proportional to $\rho_{2n} (S_x S_x^{2n-1})^\ast + \rho^\ast_{2n}
S_x S_x^{2n-1}$. The conclusions of this paragraph are well known from
studies of the spin-density-wave state of chromium.\cite{faw88}

\section{Lockin to Weak Ferromagnetism}
Consider a single-Q spin-density-wave phase characterized by a
nonzero $S_x$, as described above. A lock-in transition is
obtained by adding so-called lock-in terms proportional to $S_x^p$
and $(S_x^\ast)^p$ ($p$ is an integer) to this free energy. Since
the free energy must be invariant with respect to a translation of
the spin-density-wave by the displacement
$\tfrac{1}{2}(\mathbf{a+b+c})$, these terms are allowed only for
of $Q$ such that $Q = 2(m/p) a^\ast$ , where $m$ (as well as $p$)
is an integer. (The factor 2 in the expression $Q = 2(m/p) a^\ast$
appears because the unit cell is body centered.) Thus if $Q$ is
close to satisfying this commensurability condition for some m, it
sometimes pays the system to to adjust its $Q$ to exactly satisfy
it, so as to be able gain energy from the presence of the lock-in
terms in the free energy. If $p$ is odd, however, terms
proportional to $S_x^p$ can not exist by themselves in the free
energy since they are not invariant with respect to time reversal.
The solution to this problem for odd $p$ is to consider terms such
as $S_x^p S_{0x}$,\cite{zhu86} where $S_{0x}$ is the ferromagnetic
component of the spin density, which are invariant under time
reversal,  and since $Q = 2(m/p) a^\ast$, are also invariant under
body-centered Bravais lattice translations. Note that symmetry
requires the ferromagnetic moment to be in the same direction as
the spin-density-wave polarization vector [in this case both are
directed along the x (or a) axis]. Thus, for $p$ odd, the terms
stabilizing a lockin transition are
\begin{equation}
F_{lockin} = A_0 S_{0x}^2 + B_0 |S_x|^p cos(p\phi)S_{0x},
\label{FLockin}
\end{equation}
where $A_0 > 0$ is expected, and $S_x = |S_x|exp(i\phi)$. These
terms must be added to the original free energy given above.
Because the second term in $F_{lockin}$ is linear in $S_{0x}$, the
free energy has its minimum at a non zero value of $S_{0x}$ in the
locked in phase, and weak ferromagnetism is thereby necessarily
induced at an odd-$p$ lockin phase transition. For a continuous
lockin transition corresponding to even $p$, ferromagnetism will
not automatically be induced, and thus would not be expected to
occur simultaneously with the lockin. It should be noted that the
lockin terms of Eq.\ \ref{FLockin} exist in the commensurate phase
free energy independently of whether the transition to the
commensurate phase is first or second order.  Thus, ferromagnetism
necessarily occurs in any commensurate phase with $Q = 2(m/p)
a^\ast$ where $p$ is odd.

If $B_0 > 0$ is assumed, the minimum of $F_{lockin}$ occurs for
$p\phi = 2r\pi$ and $S_{0x} < 0$, or for $p\phi = (2r+1)\pi$ and
$S_{0x} > 0$, where r is an integer. Assuming $p = 11$, as is
appropriate for TbNi$_2$B$_2$C (since
$Q = 2(m/p)a^\ast = (6/11)a^\ast$ in the
locked in phase), and taking $r = 1, 2, ... 11$ in
$p\phi = 2r\pi$, one finds eleven distinct commensurate domains, each with
$S_{0x} < 0$.  The new commensurate unit cell has a-axis length
$11a$, and the eleven disinct domains are obtained by starting
with one of them, and then producing the others by translating
the spin-density-wave structure by the original lattice constant
$a$ ten times.  Translating each of these different commensurate
domains by the displacement $\tfrac{1}{2}(\mathbf{a+b+c})$ gives
a corresponding domain with all of the spin directions (including
the ferromagnetic component $S_{0x}$) reversed in sign.  An
incommensurate phase can be viewed as a sequence of commensurate
domains separated by domain walls, which are often
called discommensurations.\cite{mcm76}  In a continuous
incommensurate to commensurate transition, the domain walls are
swept out to the sample boundary as the temperature approaches
the transition temperature.  In a commensurate to incommensurate
transition, domain walls must be nucleated in the commensurate
phase to form the incommensurate phase.  Pinning of the domain
walls, as well as the energy barrier required to nucleate a
domain wall, contribute to hysteresis effects for both first order
and second order lockin transitions.  The lockin transition will
be first order if the interaction between domain walls is attractive,
and second order if the interaction between domain walls is
repulsive (e.g. as in \cite{jac80}).

Consider an incommensurate phase viewed as a periodic array of
discommensurations, each of which separates two commensurate domains
differing by a translation of $\tfrac{1}{2}(\mathbf{a+b+c})$.
Suppose also that the commensurate domains all correspond to a
particular wave vector $Q = (m/n)a^\ast$ with  $m$ even and $n$
odd so that each commensurate domain is ferromagnetic.  However,
from what was said above, neighboring commensurate domains have
their ferromagnetic moments in opposite directions so that the
incommensurate phase as a whole has no net ferromagnetic moment.
Applying an external magnetic field in a direction parallel or
antiparallel to the ferromagnetic moments of the domains will then
cause the domains with moments parallel to the external field
to grow (by movement of the discommensurations) at the expense of
those with moments antiparallel to the externa field, thus giving
a sort of field induced ferromagnetism to the incommensurate phase.
This effect will be inhibited by discommensuration pinning.

\section{Magnetic even-order harmonics}
Now consider a magnetic state in which there is a ferromagnetic
moment $S_{0x}$ and a spin-density-wave (which may be either
commensurate or incommensurate for the purposes of this section)
characterized by the complex amplitude $S_x$.
Spin-density-wave harmonics of (even) order $2n$ and complex
amplitude $S_{2n,x}$ are induced in the combined ferromagnetic
and spin-density-wave states by terms in the free energy given by
\begin{equation}
F_{even} =
S_{0x}[S_{2n,x}(S_x^\ast)^{2n} + S_{2n,x}^\ast S_x^{2n}].
\end{equation}
Since this contribution to the  free energy is linear in
$S_{2n,x}$, magnetic even-order harmonics are necessarily
present in a state containing both a spin-density wave
and ferromagnetism.  Conversely, a state containing both
a spin-density wave and one of its even harmonics,
necessarily contains a ferromagnetic moment.

It has been found that, in ErNi$_2$B$_2$C, even-order magnetic
harmonics are present below $T_{WFM}$.\cite{cho01,kaw02}  This
very unusual result is explained by the discussion of the
preceding paragraph.

\section{Discussion}
The model described above appears to account well for the
simultaneous continuous lockin and weakly ferromagnetic phase
transitions which occur at approximately 8K in TbNi$_2$B$_2$C. In
particular, the commensurate wave vector $Q = (6/11)a^\ast$
is of the form $Q = (m/n)a^\ast$ with $m$ even and $n$ odd.  Thus, the
continuous lockin transition is necessarily accompanied by the
presence of weak ferromagnetism.  The direction of the induced
weak ferromagnetic moment is required by symmetry to be in the
same (or opposite) direction as the polarization vector of the
spin-density wave, which is consistent with the observation
of\cite{cho96} that the induced ferromagnetic moment is in the
basal plane. If there are ferromagnetic domains present, the
lowest free energy type of domain structure is expected to have domains of
different ferromagnetic spin orientations associated with
commensurate  regions shifted relative to one another by the
displacement $\tfrac{1}{2}(\mathbf{a+b+c})$. Even harmonics have
been observed very recently \cite{kre02} both above and below the
transition temperature to weak ferromagnetism, and it would be of
interest to determine the magnetic character of these satellites
as a function of temperature. (In the model described above, even
harmonics of magnetic character should occur only in the weakly
ferromagnetic phase).  Also, \cite{kre02} has presented evidence
that the spin-density-wave phase of TbNi$_2$B$_2$C may not be
perfectly longitudinal, at least at some temperatures. A
reorientation of the spin polarization should not affect the
general arguments of this article concerning the lockin to weak
ferromagnetism.

Some features of the
weakly ferromagnetic phase observed in ErNi$_2$B$_2$C are
accounted for by the above work.  For example, the even harmonics which occur
in the ferromagnetic phase\cite{cho01,kaw02}, and which have
the unusual characteristic of being magnetic, are shown to necessarily
accompany the presence of ferromagnetism combined with a
spin-density wave.

However, the observed magnitude $Q$ of the locked in wave
vector in ErNi$_2$B$_2$C does not obviously satisfy the
requirement $Q = (m/p) a^\ast$, with $m$ even and $p$ odd, derived
above for  continuous phase transition to a locked phase that is
also ferromagnetic.  The commensurate wave vector found in
\cite{cho01} is $Q = 0.548 a^\ast = (17/31) a^\ast$, whereas the
commensurate wave vector found in \cite{kaw02,det02} is $Q = 0.550
a^\ast = (11/20) a^\ast$. It should be noted, however, that the
value $Q = (28/51) a^\ast = 0.549 a^\ast$\cite{jen02}, and the
value $Q = (16/29) a^\ast = 0.5517 a^\ast$ would both produce a
locked in weakly ferromagnetic phase as described above.  The
question now arises as to what is the significance of variations
of wave vector of the order of $\delta Q \approx 0.001$. The
transition from an incommensurate to a commensurate phase is
brought about by sweeping the domain walls between different
commensurate regions out of the crystal.  If it turns out that the
pinning of the domain walls does not allow this process to be
completed, so that there remain domain walls separated on average
by a distance $1000a$, then this will cause a change in the
observed $Q$ of order $\delta Q \approx 0.001 a^\ast$.  Clearly,
the interpretation of the locked in phase of TbNi$_2$B$_2$C, where
the commensurate wave vector involves smaller integers and the
absolute precision of its measurement is less significant, is more
straight forward than the interpretation of the locked in phase of
ErNi$_2$B$_2$C.

It is also of interest to compare the results of this article with
those of the microscopic mean field calculation of \cite{jen02}.
Ref.\ \cite{jen02} studied a number of commensurate
spin-density-wave structures having $Q$ close to $Q = 0.55 a^\ast$
for a detailed model of ErNi$_2$B$_2$C.  The commensurate
structures of that article having $Q = (6/11) a^\ast$ and $Q =
(4/7) a^\ast$ both had ferromagnetic moments, whereas
the structure with $Q = (5/9) a^\ast$ had no
ferromagnetic moment. These results are consistent with the
general rule stated above that weak ferromagnetism is necessarily
present in a commensurate phase with $Q$ of the form $Q = (m/n)
a^\ast$ where $m$ is even and $n$ is odd.  On the other hand,
Ref.\ \cite{jen02} finds a first order phase transition within the
$Q = (11/20) a^\ast$ commensurate phase from a non-ferromagnetic
to a weakly ferromagnetic state. This does not contradict the
general rule given above, since the rule does not require a
commensurate phase of this wave vector to be ferromagnetic, nor
does it prevent a particular choice of the model interactions from
yielding ferromagnetism at this wave vector. In this case,
however, there is no reason for the ferromagnetism to occur
simulataneously with the lockin (and \cite{jen02} does not argue
that it does). Finally, it is interesting to note that certain
aspects of the neutron scattering data of \cite{cho01} were
explained in \cite{jen02} by invoking a commensurate phase with $Q
= (28/51) a^\ast$. Interestingly, the symmetry arguments of this
article require that such a phase is weakly ferromagnetic. (Ref.
\cite{jen02} is silent on the question of the ferromagnetism of
their $Q = (28/51) a^\ast$ phase.)

\section*{Acknowledgements}
We thank P. C. Canfield for useful comments, and MBW thanks K. A.
Moler for drawing the interesting properties of the weakly
ferromagnetic phase of ErNi$_2$B$_2$C to his attention. Also,
support from the Canadian Institute for Advanced Research and from
the Natural Sciences and Engineering Research Council of Canada is
acknowledged.

\end{document}